\documentclass[12pt,preprint]{aastex}

\usepackage{emulateapj5}
\usepackage{epsf}

\shorttitle{The Earliest Optical Observations of GRB 030329}
\shortauthors{Torii, et al.}

\begin{document}

\title{The Earliest Optical Observations of GRB 030329}

\author{K. Torii\altaffilmark{1}, T. Kato\altaffilmark{2}, H. Yamaoka\altaffilmark{3}, T. Kohmura\altaffilmark{4}, Y. Okamoto\altaffilmark{5}, K. Ohnishi\altaffilmark{6}, K. Kadota\altaffilmark{7}, S. Yoshida\altaffilmark{8}, K. Kinugasa\altaffilmark{9}, M. Kohama\altaffilmark{1}, T. Oribe\altaffilmark{10}, and T. Kawabata\altaffilmark{11}}

\altaffiltext{1}{Cosmic Radiation Laboratory, RIKEN, 2-1, Hirosawa, Wako, Saitama 351-0198, Japan}

\altaffiltext{2}{Department of Astronomy, Faculty of Science, Kyoto University, Sakyou-ku, Kyoto 606-8502, Japan}

\altaffiltext{3}{Department of Physics, Faculty of Science, Kyushu University, Ropponmatsu, Fukuoka 810-8560, Japan}

\altaffiltext{4}{Kogakuin University, 2665-1, Nakano-machi, Hachioji, Tokyo 192-0015, Japan}

\altaffiltext{5}{Yatsugatake camera, Showa Kaken Co., 2-2-4 Nobi, Yokosuka, Kanagawa 239-0841, Japan}

\altaffiltext{6}{Nagano National College of Technology, 716 Tokuma, Nagano 381-8550, Japan}

\altaffiltext{7}{MISAO Project, 791-8, Koshikiya, Ageo, Saitama 362-0064, Japan}

\altaffiltext{8}{MISAO Project, 2-4-10-708, Tsunashima-nishi, Kohoku-ku, Yokohama, Kanagawa 223-0053, Japan}

\altaffiltext{9}{Gunma Astronomical Observatory, 6860-86, Nakayama, Takayama, Agatsuma, Gunma 377-0702, Japan}

\altaffiltext{10}{Saji Observatory, Saji Astro Park, 1071-1, Takayama, Saji-son, Yazu-gun, Tottori 689-1312, Japan}

\altaffiltext{11}{Bisei Astronomical Observatory, 1723-70 Okura, Bisei, Oda, Okayama 714-1411, Japan}

\begin{abstract}

We present the earliest optical imaging observations of GRB~030329
related to SN~2003dh. The burst was detected by the HETE-2 satellite
at 2003 March 29, 11:37:14.67 UT. Our wide-field monitoring started 97
minutes before the trigger and the burst position was continuously
observed.  We found no precursor or contemporaneous flare brighter
than $V=5.1$ ($V=5.5$) in 32~s (64~s) timescale between 10:00 and
13:00 UT.  Follow-up time series photometries started at 12:51:39 UT
(75~s after position notice through the GCN) and continued for more
than 5 hours. The afterglow was $Rc= 12.35\pm0.07$ at $t=74$~min after
burst. Its fading between 1.2 and 6.3 hours is well characterized by a
single power-law of the form $f{\rm(mJy)} = (1.99\pm0.02{\rm
(statistic)}\pm0.14{\rm (systematic)}) \times (t/1\,{\rm
day})^{-0.890\pm 0.006 {\rm (statistic)}\pm 0.010 {\rm (systematic)}}$
in $Rc$-band. No significant flux variation was detected and upper
limits are derived as $(\Delta f/f)_{\rm RMS} = 3-5$\% in minutes to
hours timescales and $(\Delta f/f)_{\rm RMS} = 35-5$\% in seconds to
minutes timescales. Such a featureless lightcurve is explained by the
smooth distribution of circumburst medium. Another explanation is that
the optical band was above the synchrotron cooling frequency where
emergent flux is insensitive to the ambient density
contrasts. Extrapolation of the afterglow lightcurve to the burst
epoch excludes the presence of an additional flare component at
$t<10$~minutes as seen in GRB~990123 and GRB~021211.

\end{abstract}
\keywords{gamma-rays: bursts --- supernovae: individual: (SN 2003dh)}

%\keywords{gamma-rays: bursts --- supernovae: individual: SN~2003dh}

\section{Introduction}

 A long ($\simeq$50~s) duration gamma-ray burst of extreme brightness,
 GRB~030329 (Vanderspek, et al. 2003), was detected by
 the three instruments, FREGATE (Atteia, et al. 2002), WXM (Shirasaki,
 et al. 2003), and SXC (Monnelly, et al. 2002) on the HETE-2 satellite
 (Ricker, et al. 2002) at 2003 March 29 11:37:14.67 UT.  The fluence
 in 3--400~keV range was $1.2\times 10^{-4}\, {\rm ergs\, cm^{-2}}$
 (Ricker, 2003). The burst direction was determined by ground analysis
 and distributed through GCN (Barthelmy, et al. 1995) at 12:50:24 UT
 (73 minutes after trigger).

 Identification of an optical afterglow was reported by Peterson \&
 Price (2003), followed by Torii (2003) and Uemura (2003).  The
 redshift of the afterglow was measured as $z=0.1685$ from high
 resolution spectroscopy with the VLT (Greiner, et al. 2003). As the
 afterglow faded, a supernova component (SN~2003dh) emerged which was
 clearly observed both photometrically and spectroscopically (Stanek,
 et al. 2003; Kawabata, et al. 2003; Hjorth, et al. 2003). A dwarf
 host galaxy with an absolute magnitude similar to the Small
 Magellanic Cloud was found with the HST observations (Fruchter, et
 al. 2003).

 In this Letter, we report the earliest phase behavior of the optical
afterglow of GRB~030329. We use a luminosity distance of 800~Mpc and
the prompt gamma-ray energy output, $E_\gamma = 7.9\times 10^{51}\,
k\, f_b\,{\rm ergs}$, where $k$ and $f_b$ are correction factors for
the observed energy range and beaming (Bloom, et al. 2003).

\section{Observations}

 Our observation log is summarized in table 1.  The sky region of
 GRB~030329 was continuously imaged by the Yatsugatake Camera.  The
 system consists of two (North and South) unguided (fixed) ultra wide
 field video cameras placed at an altitude of 1000 m in Yatsugatake,
 Japan. The sky position of the GRB was always out of the field of
 view of the northern camera. The southern camera (hereafter YC-S)
 utilized SONY XC-75 and a wide angle lens of focal length
 3.5mm (f/1.4). The field of view was $85^\circ \times
 70^\circ$. Images were continuously output in NTSC format after 8 s
 integrations, stored to frame memory (MSJ SS-10), and written to
 timelapse video tape.  Approximate read-out time of each frame was
 superimposed in the data. We further calibrated the read-out
 time of YC-N to within $\pm 4$~s by using a bright meteor.

Follow-up observations of the burst position were made with the
Automated Response Telescope (ART) at RIKEN. The two computer
controlled equatorial mounts were used for the observation. These
mounts carry a 0.20-m f/4.0 reflector (hereafter T1) and a 0.25-m
f/6.8 reflector (T2), respectively.  T1 was equipped with Apogee AP7p
camera which gave 53$\times$53 arcmin field of
view. T2 was equipped with Apogee AP6E
camera with 50$\times$50 arcmin field of view. The CCDs were
continuously read out with a constant integration times of 30-s and
60-s for T1 and T2, respectively.
%%%
 The burst position was received through the GCN (GRB Coordinates
 Network) (Barthelmy, et al. 1995) and the two telescopes
 automatically slewed to the notified position under fine
 meteorological condition. The first useful frames were acquired
 starting at 12:51:39 (75~s after notification) with T1 and 12:52:09
 with T2.
On Mar. 29, the first images obtained with T2 were matched to
 USNO-A2.0 catalog (Monet, et al. 1998) by using
 PIXY\footnote{http://www.aerith.net/misao/pixy1/index.html}. The
 optical afterglow candidate was identified as an only one
 uncataloged object in the field of view. This object was not present
 in a DSS (POSS E) frame and no corresponding
 minor planet\footnote{http://scully.harvard.edu/$^\sim$cgi/CheckMP}
 was cataloged. The information of this new transient was reported in
 a GCN circular (Torii, 2003). An earlier identification of the
 afterglow at the Siding Spring Observatory was reported by Peterson
 \& Price (2003). Between 13:41 UT and 13:43 UT, three 30-s exposures
 were obtained with ART T1 by stopping the sidereal tracking. The
 afterglow was detected as linear trails in these frames and used to
 study short time variability with 0.44~s(/pixel) resolution.

 In the city of Ageo, K. Kadota received an e-mail alert message and
observed the afterglow.  The instrument was 0.25~m Newtonian reflector
equipped with unfiltered SBIG ST-9E CCD camera.  The robotic GETS
0.25~m telescope at the Gunma Astronomical Observatory was also used
to observe the afterglow. A part of the GETS data suffered from
passages of clouds.  At the Saji Observatory, K. Oribe
acquired two $Rc$ and two $V$ frames with the 1.03-m reflector.

\section{Analyses and Results}

Analog data of the Yatsugatake Camera were analyzed after digitization
 as described below.  A single frame of 8-s integration is stored as a
 single frame of NTSC format (1/30 s) in a VHS video tape. This means
 that nominal playback of the tape shows a movie of one minute which
 corresponds to 4 hours in original time.  The tape was played back
 and read to a personal computer by using a capture card (Canopus,
 Power Capture Pro).  The data were stored as uncompressed AVI format
 movies of 640$\times$480 resolution. From these movie files,
 each frame of 8~s integration was extracted as an 8-bit FITS file.
 We created background frames as an average of many adjacent-time
 frames and subtracted the background from the raw frames. This
 procedure largely reduced the effect of noisy hot pixels.  We found
 that relatively faint ($\sim 5$~mag) stars were not steadily apparent
 in a single $8$~s integration frame. However, these stars could
 be steadily seen if we combine four consecutive frames.  This is
 probably explained by the fact that the CCD chip of YC-S is interline
 type and that it takes 32~s for a stellar image to pass a pixel
 structure by sidereal motion. We thus created stacked running mean of
 four frames, corresponding to 32~s integration. We also created
 stacked frames of $8\times 8$~s integration to see fainter
 objects. 

Figure 1 shows a stacked $8\times 8$~s frame at the burst trigger
 time.  At the position of the afterglow, no significant object was
 found.  In all the frames with 32~s (64~s) integration, a nearby
 $V=5.1$~mag ($V=5.5$~mag) star is clearly seen while fainter stars
 are not always detected.  We therefore derive an upper limit of
 $V=5.1$ ($V=5.5$) for 32~s (64~s) timescales between 10:00 and 13:00
 UT for any optical emission associated with the GRB. These results
 expand our previous report based on the same data set (Okamoto,
 Ohnishi, \& Torii 2003).

 The other data, obtained with cooled CCD cameras, were reduced in
standard ways. Figure 2 shows the first image obtained with the ART T1.
After dark subtraction and flat fielding, aperture photometry was
performed by a java-based software developed by T. Kato (ART and GETS
data), PIXY-2\footnote{http://www.aerith.net/misao/pixy/index.html}
(Ageo data), and Astroart (Saji data). Photometric analysis for the
Ageo data was performed by S. Yoshida.  The Saji data were calibrated
by using the standard stars of Landolt (1992), and the two frames in
each of $Rc$ and $V$ were combined to yield the following
measurements: $Rc=13.70\pm 0.05$ at 16:58:00 UT and $V=13.99\pm0.05$
at 17:02:45 UT.  For the ART, GETS, and Ageo data, we used a
comparison star at (RA, Dec)=(10 44 54.485, +21 34 29.80) (J2000;
USNO-A2.0 1050.06351075) as measured by Henden (2003), which has a
similar color ($V-Ic=0.839$) to the afterglow.

 For unfiltered observations, instrumental magnitudes were converted
to $Rc$ by using color terms for corresponding CCDs (Henden 2000). We
assumed that the afterglow had a constant color, $V-Ic=0.85$,
throughout the observations (Kinugasa, et al. 2003). This value is
close to a near simultaneous independent measurement ($V-Ic=0.86$ at
Mar 29.744 UT; Burenin, et al. 2003a). We find that our corrected $Rc$
magnitudes obtained from different unfiltered instruments (ART, GETS,
Ageo) agree within $0.07$~mag.  These measurements agree within
$0.07$~mag to the $Rc$-filtered Saji data and we estimate that our
absolute flux calibration is accurate at $\simeq 7$\% level.  Our
best-fit power-law function as described below predicts
$0.07-0.10$~mag brighter values compared to the simultaneous $Rc$ data
obtained at the Siding Spring Observatory (Price, et al. 2003). The
function also predicts $\simeq 0.07$~mag brighter values compared to
the first part of $Rc$ lightcurve obtained with RTT150 (Burenin, et
al. 2003b).

 Figure 3 shows the afterglow lightcurves. Among these, the data from
ART T2 have the longest time coverage with highest signal-to-noise
ratio; we discuss here on this data set.  We estimated a total error
of each measurement by RMS variations of comparison stars in 11--14
mag range.  Total errors (combination of statistical and other errors
except for a uniform shift of zero point) for T2 were estimated as
$\simeq 0.03$~mag at early part (when the afterglow was $Rc\simeq
12.5$) and $\simeq 0.05$~mag at the last part (when the afterglow was
$Rc\simeq 14.0$) in RMS. Photometric data from ART T2 are summarized in Table 2. We fit the ART T2 lightcurve with a
power-law function, $Rc = a + b\cdot {\rm log}\, (t/1\,{\rm day})$.
This function is statistically accepted with $\chi^2/d.o.f.=1.0$.  The
best-fit parameters are $a= 15.21\pm 0.01 ({\rm statistic})$ and $b
=2.22\pm0.02 (\rm statistic)$.  Flux density at $Rc$ band (658~nm) is
then derived as $f{\rm(Jy)} = (1.99\pm0.02{\rm (statistic)}\pm0.14{\rm
(systematic)}) \times 10^{-3} \times (t/1\,{\rm day})^{-0.891\pm 0.006
{\rm (statistic)} \pm 0.010 {\rm (systematic)}}$.  Statistic errors
are 90\% confidence values and the systematic error for the decay
index was estimated as the difference to the best-fit value for the
ART T1 lightcurve. We therefore conclude that deviation from a single
power-law decay was small ($(\Delta f / f)_{\rm RMS}
\stackrel{<}{_{\sim}} 3$~\% at first part and $(\Delta f / f)_{\rm
RMS} \stackrel{<}{_{\sim}} 5$~\% at last part) in minutes to hours
time scales during the observation. Uemura, et al. (2003b) reported
the presence of wavy structures in the early lightcurve. Since the
reported amplitude seems $\simeq 0.1$~mag in the time region of our
observations, the presence of such structures does not seem to be
highly inconsistent with the results presented herein.

 The three frames of ART T1 without sidereal tracking were
 investigated by constructing afterglow lightcurves with 0.44, 0.88,
 1.76, and 3.52~s time bins. Compared to a similar magnitude star, no
 significant short-term variation was found. The variation of these
 data (upper limit of the afterglow fluctuation) is derived as,
 $(\Delta f / f)_{\rm RMS}=35\, (\Delta t / {\rm 1~s})^{-1/2}\,
 [\%]$. Combining the results from ART T2 lightcurve, this relation is
 valid between seconds to minutes time scales.

\section{Discussion}

Afterglow lightcurves of gamma-ray bursts are generally described
within a framework of synchrotron radiation from external forward
shock (e.g., Wijers, Rees, \& M\'{e}sz\'{a}ros 1997) as applied for
the first afterglow of GRB~970228 (van Paradijs, et al. 1997; Costa,
et al. 1997). For an adiabatic evolution into uniform medium, the
shock radius is calculated as $r = 2.1\times 10^{17} (t / 1 {\rm
day})^{1/4} (E / E_\gamma)^{1/4} (n/10)^{-1/4}\, {\rm cm}$ (Sari,
Piran, \& Narayan 1998). For our lightcurve between 1.2 and
6.3 hours after burst, the radius corresponds to $r\simeq 0.03-0.05
(n/10)^{-1/4}$~pc.

 A single power-law decay in the optical data ($\alpha_1=-0.890$)
and a similar behavior in X-rays ($-0.9\pm 0.3$; Tiengo, et al. 2003)
suggest that the optical and X-ray bands were in the same segment of
the multi-wavelength spectrum divided by the typical synchrotron
frequency, $\nu_m$, and the synchrotron cooling frequency, $\nu_c$
(Sari, et al. 1998).  Just after the end of our observations, the
decay index suddenly steepened to $\alpha_2 = -1.22\pm0.03$ (Lipunov,
et al. 2003) or $\alpha_2 = -1.19\pm 0.01$ (Burenin, et al. 2003b). At
$t=0.57$~days, the decay index further steepened to $\alpha_3 \simeq
-1.9$ (Burenin, et al. 2003b; Garnavich, Stanek, \& Berlind 2003).

The value of $\Delta \alpha_{12}\simeq 0.3$ at $t\simeq 6$~hours is
close to $1/4$ as expected from the passage of a cooling frequency
through the observing band. In this case, the flux evolves
from $f \propto t^{-3(p-1)/4}$ to $f \propto t^{-(3p-2)/4}$ (Sari, et
al. 1998) and $p=2.19$ is derived for electron number
distribution. This interpretation is interesting, since the index $p$
is close to the universal value, $p\simeq 2.2-2.3$, expected from
particle acceleration in ultrarelativistic shock (e.g., Abraham, et
al. 2001).  However, there is difficulty in explaining the latter
larger break to $\alpha_3 \simeq -1.9$ which lasted over 1000~hours in
X-rays (Tiengo, et al. 2003). The steepening to $\alpha_3$ was
achromatic (Burenin, et al. 2003b) and could probably be interpreted
as jet break in constant density medium (e.g., Sari, Piran, \& Halpern
1999; Kumar \& Panaitescu 2000). For a jet break, the decay index
evolves from $\alpha_1=-(3p-2)/4$ to $\alpha_3=-p$.
While this interpretation suggests a
flat particle distribution, $p \simeq 1.9$ (with high energy cut-off),
the evolution of decay index is naturally reproduced.

Since this gamma-ray burst was related to the supernova (Stanek, et
al. 2003; Kawabata, et al. 2003; Hjorth, et al. 2003), significant
amount of circumstellar medium must have been present around the
progenitor (e.g., Chevalier \& Li 1999).  Massive stars (such as a
Wolf-Rayet star) blow blob-like winds and hydrodynamic instability is
also expected (e.g., Garcia-Segura, Langer, \& Mac Low 1996). The
shock interactions in such media modify the lightcurve while the
effect depends on the observed frequency range (e.g., Lazzati, et
al. 2002; Nakar, Piran, \& Granot 2003). At above the cooling
frequency, $\nu_c$, radiated flux is almost insensitive to the density
fluctuation for $p\simeq 2$. While the radiated flux sensitively
traces the density variation between $\nu_c$ and $\nu_m$.  In several
afterglows, bumpy structures (e.g., GRB~021004: Fox, et al. 2003) or
irregular fluctuations (GRB~030418: Smith, Rykoff, \& McKay 2003) were
observed at early phase. These features may be attributed to density
fluctuations in the circumburst medium and the
cooling frequency above the optical band (Lazzati, et al. 2002).

In the early afterglow of GRB~030329, the absence of significant
fluctuation may be attributed to the smooth ambient medium. For small
density contrasts, the overall hydrodynamical structure is not
modified and the radiated flux scales as $\Delta f / f \propto (\Delta
n / n)^{1/2}$ (Lazzati, et al. 2002). From our measurements of
$(\Delta f/f)_{\rm RMS} = 3-5$\%, upper limits for density contrasts
are estimated as $(\Delta n / n)_{\rm
RMS}=6-10$\%.  Another explanation is that a sign of inhomogeneities
did not emerge in the lightcurve because the optical band was above
$\nu_c$ during the observation.
 Burenin, et al. (2003b) observed the afterglow between 0.25
and 0.6 days and derived upper limits of 10--1\% for flux
fluctuation on 0.1--1000~s time scales. Therefore,  the
current discussions apply for a broader time range between 1.2 hours
and 0.6 days after burst.

 If we simply extrapolate the afterglow lightcurve toward the burst
epoch, expected magnitudes are $Rc=6.5$ and $Rc=8.7$ at $t=10$~s and
$t=100$~s after trigger, respectively. Our contemporaneous upper
limits of $V=5.1-5.5$, as well as similar limits from contemporaneous
photography (Sasaki, et al. 2003), are above these estimates. In some
gamma-ray bursts (GRB 990123: Akerlof, et al. 1999; GRB 021211:
Wozniak, et al. 2002; Park, Williams, \& Barthelmy 2002; Li, et
al. 2003a), bright and rapidly fading ($f\propto t^{-2}$) optical
flare is detected. This component dominates the ordinary afterglow
within 10~minutes after burst and considered to originate from reverse
shock (e.g., Kobayashi 2000). Our contemporaneous upper limit excludes
the presence of such a flare component, as well as bright precursor
emission.

In summary, the early optical afterglow of GRB~030329 is well
 characterized by a single power-law decay with index
 $\alpha_1=-0.890$. The absence of significant flux fluctuation in our
 lightcurve suggests that the circumburst medium was smoothly distributed
 or the optical band was above the synchrotron cooling frequency
 during the observation.

\acknowledgments

 This work is partly supported by a grant-in-aid
 15740129 (Torii), 13640239 and 15037205 (Kato), and 14740131
 (Yamaoka) from the Japanese Ministry of Education, Culture, Sports,
 Science and Technology.

%References

\clearpage

\begin{figure}
\epsscale{0.5}\plotone{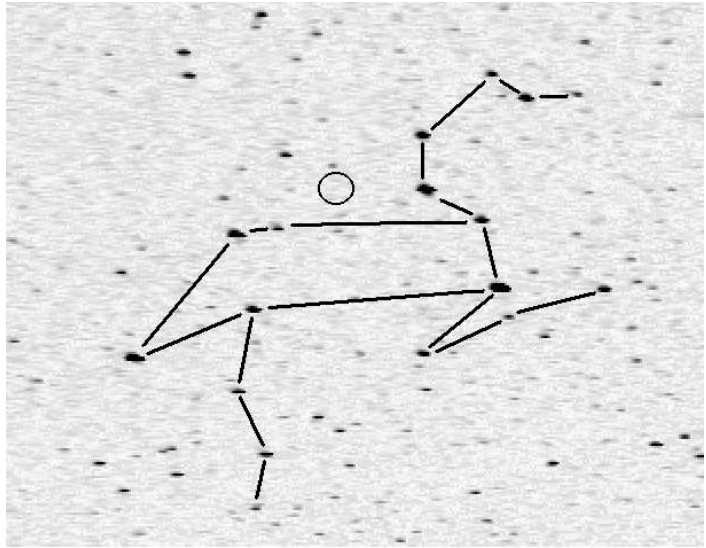}
\caption{A wide field YC-S image around the constellation of Leo at the burst trigger time\label{fig1}. This image is a stack of 8$\times$8~s frames starting at 11:36:42 UT (33~s before trigger time). A circle shows the position of the afterglow where no significant object is apparent.}
\end{figure}

\begin{figure}
\epsscale{0.5}\plotone{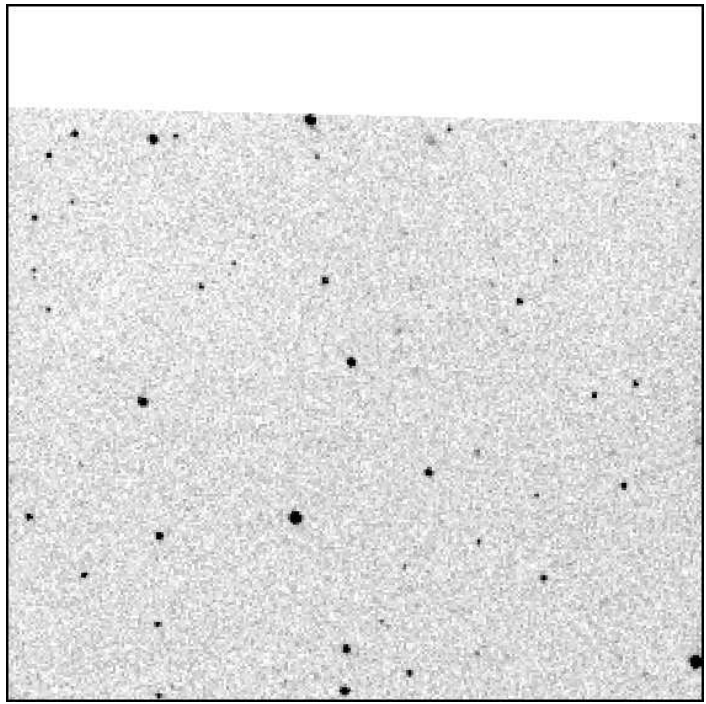}
\caption{The first 30~s exposure (mean epoch 12:52:54 UT) by ART T1. 30$'\times$30$'$ field centered at the afterglow is shown. North is up and east to the left.\label{fig2}}
\end{figure}
%\clearpage

\begin{figure}
\epsscale{0.75}
\rotatebox{-90}{
\plotone{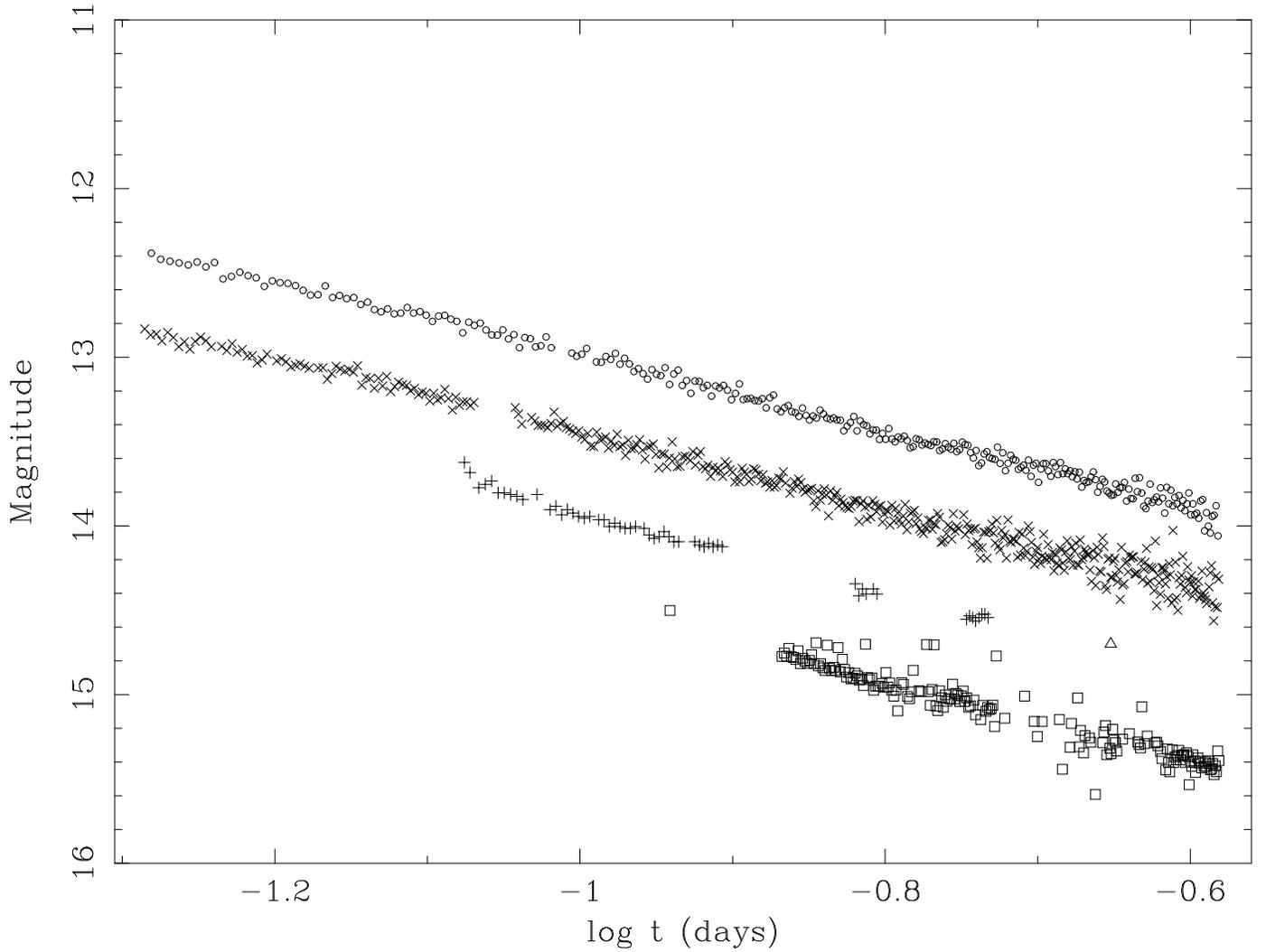}}
\caption{From top to bottom, ART T2 (circles), ART T1 (x marks), Ageo
(plus marks), Saji (triangle), and GETS (squares) lightcurves are
shown.  The latter four data sets are shifted by +0.5, +1.0, +1.0, and
+1.5 magnitudes. Total errors for T2 data are $0.03$~mag at the first
part and $0.05$~mag at the last part. For clear
presentation of ART T1 and Ageo lightcurves, a bright comparison star
with high signal to noise ratio, USNO A2.0 1050.06353017, is used
here.\label{fig3}}
\end{figure}

\clearpage

\begin{deluxetable}{lllllrrl}
\tabletypesize{\tiny}
\tablecaption{Observation Log. \label{tbl-1}}
\tablewidth{0pt}
\tablehead{
\colhead{Site} & \colhead{Instruments}   &   \colhead{Filter} &
\colhead{Start Time (UT)} &
\colhead{End Time (UT)}  & \colhead{Integration [s]} & \colhead{Number of frames} &
\colhead{Observer / PI}     
}
\startdata
Yatsugatake & YC-S & IR-cut & 10:00:00 & 13:00:00 & 8 & 1350 &  Okamoto \\
Wako & ART 0.20 m + AP7p & No & 12:51:39 & 17:55:00 &30  & 437 &  Torii \\
Wako & ART 0.25 m + AP6E & No &12:52:09 & 17:54:52 &60 & 277 &  Torii \\
Ageo & 0.25m + ST-9E & No & 13:38:12 & 16:03:50 & 4, 60 &59 & Kadota \\
Gunma & GETS 0.25m + AP7p & No & 14:21:55 & 17:55:06 & 30 & 181 & Kinugasa \\%-1579
Saji      & 1.03 m & $Rc$, $V$ & 16:54:30 & 17:05:00 & 120  & 4 &  Oribe \\
\enddata
\end{deluxetable}

\begin{deluxetable}{lll}
\tabletypesize{\tiny}
\tablecaption{Photometric data obtained by ART T2. \label{tbl-2}}
\tablewidth{0pt}
\tablehead{
\colhead{Mean Epoch $^{\rm a}$} & \colhead{$Rc$ mag} & \colhead{Mag error (1 $\sigma$)}
}
\startdata
0.052365 & 12.383 & 0.025 \\
0.053106 & 12.419 & 0.026 \\
0.053858 & 12.430 & 0.026 \\
0.054610 & 12.441 & 0.026 \\
0.055363 & 12.453 & 0.026 \\
\enddata

$^{\rm a}$ days after burst trigger at 2003 March 29, 11:37:14.67 UT.

Only the first 5 rows are tabulated here. The other data are available in electric format (machine readable table).

\end{deluxetable}

\end{document}